\documentclass{PoS}
\usepackage{amsmath}

\title{Observational Aspects of an Inhomogeneous Cosmology}

\ShortTitle{Observational Aspects of an Inhomogeneous Cosmology}

\author{\speaker{Christoph Saulder}\\
        European Southern Observatory, Chile \& Department of Astrophysics, University of Vienna\\
        E-mail: \email{csaulder@eso.org}}

\author{Steffen Mieske\\
        European Southern Observatory, Chile\\
        E-mail: \email{smieske@eso.org}}
        
\author{Werner W. Zeilinger\\
        Department of Astrophysics, University of Vienna\\
        E-mail: \email{werner.zeilinger@univie.ac.at}}

\abstract{One of the biggest mysteries in cosmology is Dark Energy, which is required to explain the accelerated expansion of the universe within the standard model. But maybe one can explain the observations without introducing new physics, by simply taking one step back and re-examining one of the basic concepts of cosmology, homogeneity. In standard cosmology, it is assumed that the universe is homogeneous, but this is not true at small scales (<200 Mpc). Since general relativity, which is the basis of modern cosmology, is a non-linear theory, one can expect some backreactions in the case of an inhomogeneous matter distribution. Estimates of the magnitude of these backreactions (feedback) range from insignificant to being perfectly able to explain the accelerated expansion of the universe. In the end, the only way to be sure is to test predictions of inhomogeneous cosmological theories, such as timescape cosmology, against observational data. If these theories provide a valid description of the universe, one expects aside other effects, that there is a dependence of the Hubble parameter on the line of sight matter distribution. The redshift of a galaxy, which is located at a certain distance, is expected to be smaller if the environment in the line of sight is mainly high density (clusters), rather than mainly low density environment (voids). Here we present a test for this prediction using redshifts and fundamental plane distances of elliptical galaxies obtained from SDSS DR8 data. In order to get solid statistics, which can handle the uncertainties in the distance estimate and the natural scatter due to peculiar motions, one has to systematically study a very large number of galaxies. Therefore, the SDSS forms a perfect basis for testing timescape cosmology and similar theories. The preliminary results of this cosmological test are shown in this contribution.}

\FullConference{VIII International Workshop on the Dark Side of the Universe,\\
		June 10-15, 2012\\
		Rio de Janeiro, Brazil}

\begin{document}
\section{Timescape cosmology}
\begin{figure}[ht]
\begin{center}
\includegraphics[width=8cm]{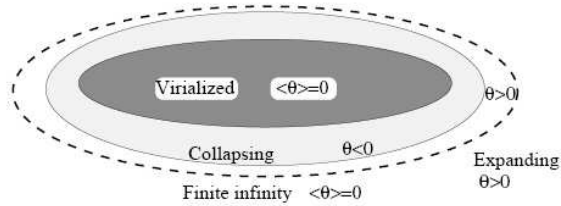}\\ 
\caption{A schematic illustration of the concept of finite infinity (by David Wiltshire \cite{Wiltshire2007a}).}
\label{finiteinfinity}
\end{center}
\end{figure}

Inhomogeneous cosmology has been around since the days of Tolman \cite{Tolman1934} and Bondi \cite{Bondi1947}, but for a very long time it was a rather quiet and exotic topic. During the last 15 years significant advances were made on this field, mainly due to the work of groups around Buchert \cite{Buchert1997a,Buchert2000b,Buchert2000c,Buchert2002,Buchert2003a,Buchert2003b,Buchert2009,BuchertRasanen2011} , R{\"a}s{\"a}nen \cite{Rasanen2004b,Rasanen2006b,BuchertRasanen2011,Rasanen2011}, Wiltshire  \cite{Wiltshire2007a,Wiltshire2007b,Wiltshire2008b,Wiltshire2009c,Wiltshire2011a}
 and others. The basic assumption is that since general relativity is a non-linear theory, inhomogeneities like voids and cluster can cause some backreactions (feedback) on cosmological parameters, which may explain the observed accelerated expansion of the universe. Buchert constructed a scheme \cite{Buchert2000b}, which is based on perturbation theory and general relativity, and it considers the inhomogeneities' influence on the average properties of cosmological parameters. In the simple case of a general relativistic dust, the equations, which describe the cosmic expansion, have to be modified to the Buchert's scheme: 
\begin{equation}
\begin{split}
\begin{array}{lcr} 3 \left(\frac{\dot{\bar{a}}}{\bar{a}}\right)^{2}=8 \pi G \left\langle\rho\right\rangle -\frac{1}{2}\left\langle R \right\rangle - \frac{1}{2} Q & \,\,\,\,\,\,\,\,\,\, &
3 \frac{\ddot{\bar{a}}}{\bar{a}}=-4 \pi G \left\langle\rho\right\rangle + Q \\
\partial_{t} \left\langle\rho\right\rangle + \frac{\ddot{\bar{a}}}{\bar{a}} \left\langle\rho\right\rangle = 0 & \,\,\,\,\,\,\,\,\,\, &
Q = \frac{2}{3} \left\langle \left( \theta - \left\langle \theta \right\rangle\right)^{2} \right\rangle - 2 \left\langle \sigma \right\rangle^{2}\end{array}
\end{split}
\end{equation}
The backreaction $Q$ is defined by the expansion $\theta$ and the shear $\sigma$. $\bar{a}$ is the scaling parameter of the universe, $\left\langle R \right\rangle$ the average spatial curvature, $\left\langle\rho\right\rangle$ the average energy density and $G$ the gravitational constant. But the acceleration of the universe's expansion cannot be fully understood in a simple pertubative approach alone \cite{Rasanen2006b,Kolb2006,Ishibashi2006}. One of the most advanced conceptions of an inhomogeneous cosmology, which can mimic dark energy, was created by Wiltshire \cite{Wiltshire2007a} and it is called ''timescape cosmology''. He uses a simple two-phase model consisting of a fractal bubble of empty voids and dense walls (clusters and filaments). Both regions are separated by the finite infinity boundary (see Fig \ref{finiteinfinity}), which encloses gravitationally bound regions and disconnects them from the freely expanding voids. 
In this model, a backreaction also causes significant differences in the time flow, due to effects of quasilocal gravitational energy: the universe in the middle of a void is older than in the centre of a cluster. Due to this effect, this specific model of inhomogeneous cosmology is also called timescape cosmology. As a consequence of the importance of the local geometry in this model, the Hubble flow is not uniform any more and the empty voids expand faster than the dense walls. At large scales, these different expansion rates will lead to the signature of an overall accelerated expansion of the universe, because in timescape cosmology the fraction of the volume occupied by voids constantly increases with time. According to Wiltshire \cite{Wiltshire2007a}, the dynamics of this fractal bubble model can be described by following equations: 
\begin{equation}
\begin{split}
\left(\frac{\dot{\bar{a}}}{\bar{a}}\right)^{2} + \frac{\dot{f}_{v}^{2}}{9 f_{v} \left(1 - f_{v} \right)} - \frac{\alpha^{2} f_{v}^{\frac{1}{3}}}{\bar{a}^{2}} = \frac{8 \pi G}{3} \bar{\rho}_{0} \frac{\bar{a}_{0}^{3}}{\bar{a}^{3}} \\
\ddot{f}_{v} + \frac{\dot{f}_{v}^{2}\left(2 f_{v} - 1\right)}{2 f_{v} \left(1 - f_{v}\right)} + 3 \frac{\dot{\bar{a}}}{\bar{a}} \dot{f}_{v} - \frac{3 \alpha^{2}  f_{v}^{\frac{1}{3}} \left(1 - f_{v}\right) }{2 \bar{a}^{2}} = 0
\end{split}
\end{equation}
The variable $f_{v}$ denotes the volume fraction of voids in the universe, which is of course time dependent and $\bar{\rho}_{0}$ is the true critical density \cite{Wiltshire2007a,Buchert2002,Buchert2003b}. Recently there have been several papers \cite{Bose2011,Clarkson2011,Umeh2011,Bull2012,Clifton2012,Paranjape2008,Bochner2012,Wiltshire2011a,BuchertRasanen2011}
, which show that the magnitude and importance of these backreactions is still a topic of hot discussion. Timescape cosmology and similar inhomogeneous cosmologies may provide possible solutions for the dark energy problem, but the estimates of the magnitude of backreaction from voids and their influence on the expansion of the universe range from negligible to extremely important \cite{Mattsson2011,Kwan2009,Paranjape2008,Wiltshire2007a}. Therefore, observational tests are essential for the ongoing debate.
\section{Predictions of the theory}
\begin{figure}[ht]
\begin{center}
\includegraphics[width=10cm]{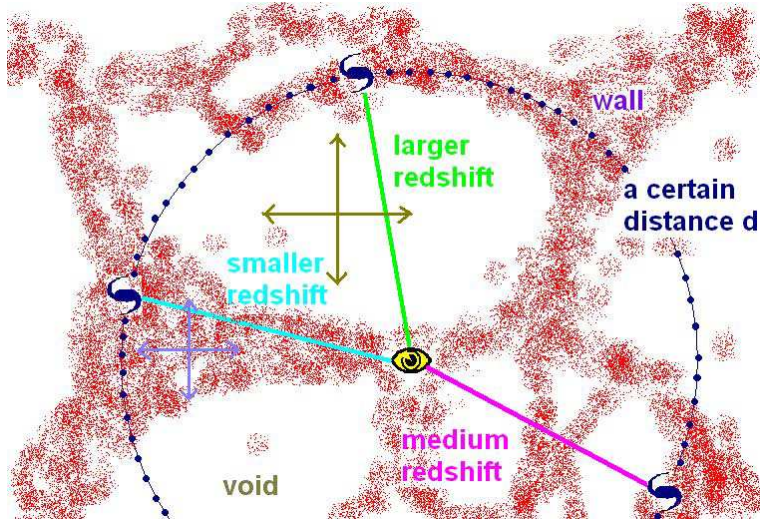}\\ 
\caption{The measured redshift at a fixed distance depends on matter distribution in the line of sight.}
\label{skizze}
\end{center}
\end{figure}

There are several predictions of timescape cosmology, which can be used as potential tests. Most of them are extremely difficult and not possible with today's technology or leave quite some space for interpretation and therefore, they cannot produce striking evidence neither for nor against the theory. Here we focus on a very direct test which was proposed \cite{Schwarz2010,Wiltshire2009c}, namely measuring the different expansion rates of voids and walls directly. Those should differ by about 17 to 22\% \cite{Wiltshire2011a}, in order to fully explain the observed accelerated expansion with timescape cosmology: The Hubble parameter is larger, if the foreground is void dominated, rather than wall dominated (for a better illustration of this feature see Fig. \ref{skizze}).
This test requires: 1. redshift data, 2. a redshift independent distance indicator and 3. a model of the matter distribution in the line of sight. While performing this test one might encounter potential problems like uncertainties in the distance measurement, peculiar motions of the galaxies and difficulties in mass estimates for matter distribution. Due to the statistical nature of these problems, one can handle them quite well using a large homogeneous sample.
\section{Testing the predictions}
We use data from the Sloan Digital Sky Survey (SDSS) DR8 \cite{SDSS} in order to perform our test. We take redshifts, central velocity dispersions, the different apparent model magnitudes in the 5 SDSS filters and the corresponding effective radii of these models from the SDSS DR8 database. Furthermore, we make use of third party information, which is also implemented in the SDSS database such as the extinction map of Schlegel \cite{Schlegelmaps} and Galaxy classification from the citizen-science project GalaxyZoo \cite{GalaxyZoo,GalaxyZoo_data}, which is based on SDSS. In addition to that, we also use masses from the SDSS-based catalogue of groups and clusters by Yang et al. \cite{Yang2007} and the new high-quality K-correction by Chilingarian et al. \cite{Chilingarian2010}.
\subsection{Calibrate the fundamental plane}
\begin{figure}[ht]
\begin{center}
\includegraphics[width=10cm]{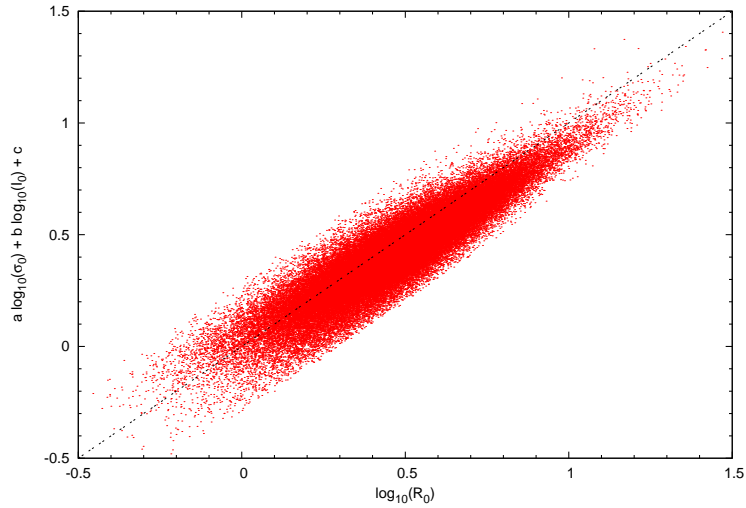}\\ 
\caption{The fundamental plane of elliptical galaxies fitted for the SDSS r-band using 90 000 galaxies.}
\label{fp_r_poster}
\end{center}
\end{figure}

The fundamental plane of elliptical galaxies is an empirical relation between the effective radius $R_{0}$, the mean surface brightness $-2.5 \cdot \textrm{log}\left(I_{0}\right)$ and the central velocity dispersion $\sigma_{0}$ of these galaxies, which can be used as redshift independent distance indicator. 
\begin{equation}
R_{0} = a \cdot \textrm{log}(I_{0}) + b \cdot \textrm{log}(\sigma_{0}) + c
\end{equation}
We calibrate this relation, in a similiar manner as Bernardi et al. \cite{Bernardi2003} did, but using more than 90 000 elliptical galaxies from SDSS, which were classified by GalaxyZoo \cite{GalaxyZoo,GalaxyZoo_data} and by applying some additional constraints to avoid misclassifications. One can derive all three parameters of the fundamental plane directly from observables which are already in SDSS data only using the Schlegel extinction maps \cite{Schlegelmaps} and the Chiligarian K-corrections \cite{Chilingarian2010} for corrections. The resulting fit for r-band data can be seen in Fig. \ref{fp_r_poster} for which we obtain a root mean square of about 10\%. The results will be published in an upcoming paper (Saulder et al. 2012, in preparation). We will use the fundamental plane to calculate distances to a quality selected subsample of about 10 000 elliptical galaxies.
\subsection{The foreground model}
\begin{figure}[ht]
\begin{center}
\includegraphics[width=10cm]{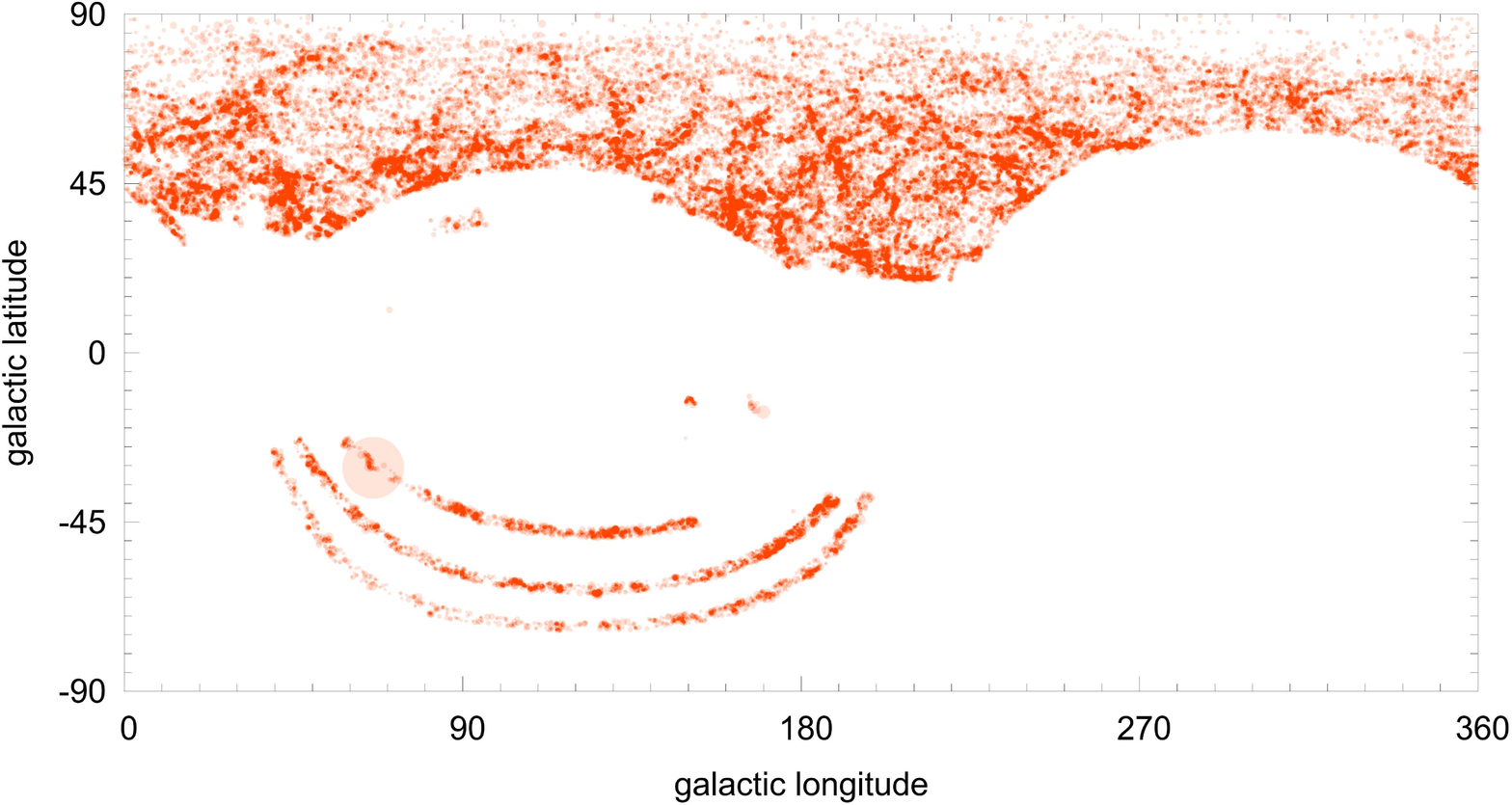}\\ 
\caption{A part of the foreground model between 100 and 150 $h^{-1}$ Mpc. One can also see the sky coverage of SDSS here.}
\label{foreground}
\end{center}
\end{figure}

We use data of more than 350 000 galaxies to model the foreground. The masses of galaxy groups and clusters are taken from the Yang catalogue \cite{Yang2007} and since it is only based on DR4, which has a smaller sky coverage than DR8, we extended it using mass-light ratios for all missing objects. We plan to do this more sophisticated in the future and create a similar (using the same methods) but larger catalogue as Yang et al. We calculate the radii of homogeneous spheres with renormalized critical density (finite infinity regions) around the clusters and galaxies in our foreground model. The distances for the objects in the foreground model are simply estimated using a redshift-distance relation. A part of our foreground model can be see in Fig. \ref{foreground}.
\subsection{Testing timescape cosmology}
For the final analysis, we use redshifts and fundamental plane distances to calculate ''individual Hubble parameters'' for every galaxy in the sample. Furthermore, the fraction of the line of sight which is in wall environment (inside a finite infinity boundary) is calculated using the foreground model. This can be done using simple geometry (intersecting straight lines with spheres) and interval nesting, but it has to be done more than 10 000 $\times$ 350 000 times. Consequently, this requires a lot of computational power for which we use the ViennaAstroCluster. In a final step, one has to put the fraction of the line of sight inside wall environment in relation to the ''individual Hubble parameters''.
\section{Preliminary results}
\begin{figure}[ht]
\begin{center}
\includegraphics[width=10cm]{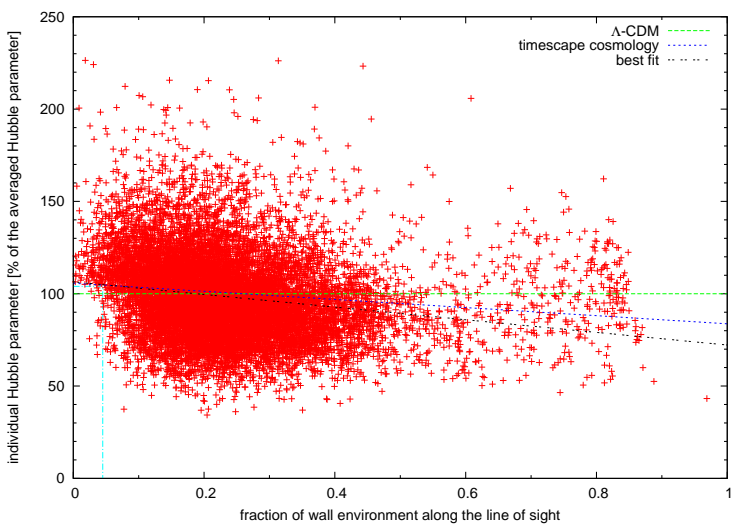}\\ 
\caption{This plot shows the dependence of the Hubble parameter on the foreground matter distribution.}
\label{testtimescape}
\end{center}
\end{figure}

Our preliminary analysis yields systematically larger Hubble parameters for low density environment (voids) in the line of sight (see Fig. \ref{testtimescape}). The distribution is not as smooth as may be expected, given the dearth of galaxies for void foreground and below average Hubble parameter. This is still a matter of concern for us in this analysis. It might be due to yet unknown biases or unknown systematic effects or maybe further improvements in the foreground model are necessary (Saulder et al. 2013b, in preparation). Concerning Fig. \ref{testtimescape}: The $\Lambda$-CDM estimate of no dependence on the line of sight environment at all is too na\"{\i}ve since it does not take into account coherent infall into clusters, which creates a similar effect of yet unknown magnitude (a comparison with large cosmological N-body simulations will be included in an upcoming paper (Saulder et al. 2013b, in preparation)). Furthermore, it should be noted that any fit to a distribution with such a scatter strongly depends on the fitting method (for example using a binned fit instead of the least square method for which the result is shown in Figure \ref{testtimescape}, one can get a very different dependence) and therefore, we cannot yet conclude any clear evidence although the preliminary data looks promising. 
We also want to point out that this project creates quite some additional science output on its way, because we obtain new fits for the fundamental plane (Saulder et. al, 2013a in preparation) and it yields lots of data on peculiar velocities of galaxies and on the large scale structure of the local universe aside from testing timescape cosmology.
\section*{Acknowledgements}
CS thanks the Department of Astrophysics of the University of Vienna for funding his participation at the conference. CS is supported by an ESO-studentship.

\end{document}